\documentclass{article}

\usepackage{amsmath, amssymb, latexsym, amsfonts}
\usepackage{epsf}

\def\sect#1{\section{#1}\setcounter{equation}{0}}

\def\q{\boldsymbol{q}}
\def\w{\boldsymbol{w}}

\begin{document}

\centerline{}
\vskip0.5cm


\centerline{} \vskip0.5cm \centerline{\Large \bf
Discrete Mechanics Based on Finite Element
Methods} \vskip0.7cm \centerline{Jing-Bo
Chen$^{\dag}$, \,\, Han-Ying Guo$^{\dag}$ \,\,
 and \,\, Ke Wu$^{\ddag\dag}$ }
\vskip6pt \centerline{\small $^{\dag}$Institute
of Theoretical Physics, Chinese Academy of
Sciences} \vskip3pt \centerline{\small P.O. Box
2735, Beijing 100080, P.R. China} \vskip6pt
\centerline{\small $^{\ddag}$ Department of
Mathematics, Capital Normal University} \vskip3pt
\centerline{\small  Beijing 100037, P.R. China}
\vskip3pt \centerline{\small
\texttt{<chenjb><hyguo><wuke>@itp.ac.cn}}\vskip
1cm



\begin{abstract}
Discrete Mechanics based on finite element methods is presented
in this paper.  We also explore the relationship between this discrete
mechanics and Veselov discrete mechanics. High order discretizations
are constructed in terms of high order interpolations.

\end{abstract}



\sect{Introduction}

We begin by recalling the variational principle
for Lagrange mechanics. Suppose $Q$ denotes the
configuration space with coordinates $q^{i}$ and
$TQ$ the tangent bundle with coordinates $(q^{i},
\dot{q}^{i})$,\,\, $i=1, 2, \cdots, n$. In the
following, we will work with the extended
configuration space $R\times Q$ with coordinates
$(t, q^{i})$ and the extended tangent bundle
$R\times TQ$ with coordinates $(t, q^{i},
\dot{q}^{i})$. Here $t$ denotes the time
[\ref{a1}].

Consider a Lagrangian
$L:\,\,R\times TQ  \to R$. The corresponding action functional is defined by
\begin{align}
    S((t, q^{i}(t)))=\int_{a}^{b}L(t, q^{i}(t), \dot{q}^{i}(t))\,dt, \label{1.1}
\end{align}
where $q^{i}(t)\in \mathcal{C}^{2}([a, b])$,
which is the set of all $C^{2}$ curves in $Q$
defined on $[a, b]$.

The variational principle seeks the curves
$q^{i}(t)$ in $\mathcal{C}^{2}([a, b])$, for
which the action functional $S$ is stationary
under variations of $q^{i}(t)$ with fixed
endpoints.

Consider a vector field on $R\times Q$
\begin{align}
 V=\xi(t, q^{j})\frac{\partial}{\partial t}
    +\sum_{i=1}^{n}\phi^{i}(t, q^{j})\frac{\partial}{\partial q^{i}}.  \label{1.2}
\end{align}
Let $F^{\epsilon}$ be the flow of $V$.  For $(t, q^{j})\in R\times Q$, we have
$F^{\epsilon}(t, q^{j})=(\tilde{t}, \tilde{q}^{i})$.
\begin{align}
  &\tilde{t}=f(\epsilon, t, q^{j}),\label{1.3}\\
  &\tilde{q}^{i}=g^{i}(\epsilon, t, q^{j}),  \label{1.4}
\end{align}
where
\begin{align}
\begin{split}
   &\left.\frac{d}{d\epsilon}\right|_{\epsilon=0}f(\epsilon, t, q^{j})
   =\xi(t, q^{j}),\\
   &\left.\frac{d}{d\epsilon}\right|_{\epsilon=0}g^{i}
      (\epsilon, t, q^{j})=\phi^{i}(t, q^{j}).
\end{split} \label{1.5}
\end{align}
 The transformation (\ref{1.3}-\ref{1.4}) transforms a curve
$q^{j}(t)$ in $Q$ into another curve $\tilde{q}^{i}(\epsilon, \tilde{t})$ in $Q$
determined by
\begin{align*}
  &\tilde{t}=f(\epsilon, t, q^{j}(t)), \\
  &\tilde{q}^{i}=g^{i}(\epsilon, t, q^{j}(t)).
\end{align*}

Before calculating the variation of $S$, we
consider the first order prolongation of $V$
\begin{align}
 \text{pr}^{1}V=\xi(t, q^{j})\frac{\partial}{\partial t}+
                \sum_{i=1}^{n}\phi^{i}(t, q^{j})\frac{\partial}{\partial q^{i}}
      + \sum_{i=1}^{n}\alpha^{i}(t, q^{j}, \dot{q}^{j})
        \frac{\partial}{\partial \dot{q}^{i}},  \label{1.6}
\end{align}
where $\text{pr}^{1}V$ denote the first order prolongation
of $V$ and
\begin{align}
      \alpha^{i}(t, q^{j}, \dot{q}^{j})=D_{t}\phi^{i}(t, q^{j})
   -\dot{q}^{i}D_{t}\xi(t, q^{j}), \label{1.7}
\end{align}
where $D_{t}$ denotes the total derivative [\ref{o1}].

Now we calculate the variation of $S$ directly
\begin{align}
\delta S=&\left.\frac{d}{d\epsilon}\right|_{\epsilon=0}
    S((\tilde{t}, \tilde{q}^{i}(\epsilon, \tilde{t})))
           \notag\\
         =&\left.\frac{d}{d\epsilon}\right|_{\epsilon=0}\int_{\tilde{a}}^{\tilde{b}}
          L\left(\tilde{t}, \tilde{q}^{i}(\epsilon, \tilde{t}), \frac{d}{d\tilde{t}}
          \tilde{q}^{i}(\epsilon, \tilde{t})\right)\,d\tilde{t}\notag\\
         =&\int_{a}^{b}\left[\left(\frac{\partial L}{\partial t}+\frac{d}{dt}\left(
           \frac{\partial L}{\partial \dot{q}^{i}}\dot{q}^{i}-L\right)\right)\xi+\left(
          \frac{\partial L}{\partial q^{i}}-\frac{d}{dt}\frac{\partial L}{\partial \dot{q}^{i}}
          \right)\phi^{i}\right]\,dt\notag\\
          &\quad +\left.\left[\left(L-\frac{\partial L}{\partial \dot{q}^{i}}
           \dot{q}^{i}\right)\xi
           +\frac{\partial L}{\partial \dot{q}^{i}}\phi^{i}\right]\right|_{a}^{b}.
         \label{1.8}
\end{align}

If $\xi(a, q^{j}(a))=\xi(b, q^{j}(b))=0$ and
$\phi^{i}(a, q^{j}(a))=\phi^{i}(b, q^{j}(b))=0$,
 the requirement of $\delta S
=0$ yields the energy evolution equation
\begin{align}
\frac{\partial L}{\partial t}+\frac{d}{dt}\left(
           \frac{\partial L}{\partial \dot{q}^{i}}\dot{q}^{i}-L\right)=0
\label{1.9}
\end{align}
from  $\xi$ and
the Euler-Lagrange equation
\begin{align}
 \frac{\partial L}{\partial q^{i}}-\frac{d}{dt}
          \frac{\partial L}{\partial \dot{q}^{i}}=0 \label{1.10}
\end{align}
from  $\phi^{i}$.{\footnote{Note that Hamilton's
variational principle holds in fact for the
vertical variation, while the horizontal
variation gives rise to the conservation relation
between the Euler-Lagrange equation and energy
conservation, see, for example, [\ref{gw}]. Here,
in order to transfer to discrete case more
directly, this variational requirement is
employed.}}

If $L$ does not depend on $t$ explicitly, i.e.,
$L$ is conservative and  $\frac{\partial
L}{\partial t}=0$, then (\ref{1.9}) becomes the
energy conservation law
\begin{align}
\frac{d}{dt}\left(
           \frac{\partial L}{\partial \dot{q}^{i}}\dot{q}^{i}-L\right)=0.
\label{1.11}
\end{align}
\noindent
If we drop the requirement
\begin{align*}
 &\xi(a, q^{j}(a))=\xi(b, q^{j}(b))=0,\\
 &\phi^{i}(a, q^{j}(a))=\phi^{i}(b, q^{j}(b))=0,
\end{align*}
we can define from the second term in
(\ref{1.8}) the extended Lagrangian one form
\begin{align}
\theta_{L}=\left(L-\frac{\partial L}{\partial \dot{q}^{i}}
           \dot{q}^{i}\right)dt +\frac{\partial L}{\partial \dot{q}^{i}}dq^{i} \label{1.12}.
\end{align}
 We define
 the extended Lagrangian two form to be
 $$\omega_{L}=d\theta_{L}.$$
Suppose $g^{i}(t, v_{\q})$ is a solution of (\ref{1.10}) depending on the initial condition
$v_{\q}\in TQ$. Restricting $\tilde{q}^{i}(\epsilon, \tilde{t})$ to the solution space
of (\ref{1.10}) and using the same method in [\ref{m1}], we can prove
\begin{align}
(\text{pr}^{1}g)^{*}\omega_{L}=\omega_{L},
\label{1.13}
\end{align}
where $\text{pr}^{1}g^{i}(s, v_{\q}))=(s,
g^{i}(s, v_{\q}), \frac{d}{ds}g^{i}(s, v_{\q}))$
denotes the first order prolongation of $g^{i}(s,
v_{\q})$.

By defining a discrete Lagrangian with fixed
steps and constructing a discrete action, Veselov
develops a kind of discrete mechanics [\ref{v1},
\ref{v2}]. See also [\ref{m3}, \ref{w1}]. The
resulting discrete Euler-Lagrange equation
(variational integrator) preserves a discrete
Lagrange two form. However the discrete versions
of (\ref{1.9}) and (\ref{1.11}) are not reflected
in Veselov discrete mechanics.
 Using variable steps and a defined discrete energy, Kane, Marsden and Ortiz obtain
symplectic-energy integrators [\ref{k1}]. Chen,
Guo and Wu develop a discrete total variation and
prove the symplectic-energy integrators preserve
a discrete extended Lagrange two form [\ref{c1}].
Guo and Wu also present the difference
variational approach with variable step-lengths
[\ref{gw}].

The connection between Veselov discrete mechanics
and finite element methods was first suggested in
[\ref{m1}]. Symplectic and multisymplectic
structures in simple finite element methods are
explored in [\ref{g1}]. As is known, it is
difficult to obtain high order discretizations in
Veselov discrete mechanics. We find that the
finite element methods provide a natural way to
obtain a kind of discrete Lagrangian, including
those of high order. The purpose of this paper is
to develop such kind of discrete mechanics based
on finite element methods.

This paper is organized as follows. In the next section, we will present the
discrete mechanics based on finite element methods.
 Section 3 is devoted to discussing the role of time steps.
 We finish this paper by drawing some conclusions in Section 4.



\sect{Discrete mechanics based on finite element methods}
\subsection{Generalized Veselov discrete mechanics}

We first present Veselov discrete mechanics  in a
more general formulation that allows for variable
time steps. In this generalized Veselov discrete
mechanics, $M\times M$ is used for the discrete
version of  $R\times TQ$. Here $M=R\times Q$. A
point $( t_{k}, q_{k};  t_{k+1}, q_{k+1})\in
M\times M$ corresponds to a tangent vector
$\frac{q_{k+1}-q_{k}}{t_{k+1}-t_{k}}$. Here for
brevity of notations, we restrict ourselves to
one-dimensional configuration space $Q$. A
discrete Lagrangian is defined to be
$\mathbb{L}:\,M\times M\to R$ and the
corresponding action to be
\begin{align}
 \mathbb{S}=\sum_{k=0}^{N-1}\mathbb{L}(t_{k}, q_{k}, t_{k+1}, q_{k+1})
   (t_{k+1}-t_{k}). \label{2.1}
\end{align}
The discrete variational principle  is to extremize $\mathbb{S}$
for variations of  $q_{k}$  holding
the endpoints $ q_{0}$ and $ q_{N}$ fixed. In this discrete variational principle,
$t_{k}$ play the role of  parameters (They can also play the role of  variables, see
next section). This discrete variational
principle  determines a discrete
flow $\Phi:\,M\times M\to M\times M$ by
\begin{align}
   \Phi(t_{k-1}, q_{k-1}, t_{k}, q_{k})=(t_{k}, q_{k},t_{k+1}, q_{k+1}). \label{2.2}
\end{align}
Here $ q_{k+1}$ is found from the following discrete Euler-Lagrange equation
(variational integrator)
\begin{align}
 &(t_{k+1}-t_{k})D_{2}\mathbb{L}(t_{k}, q_{k}, t_{k+1}, q_{k+1})
  +(t_{k}-t_{k-1})D_{4}\mathbb{L}(t_{k-1}, q_{k-1}, t_{k}, q_{k})=0
 \label{2.3}
\end{align}
for all $k\in\{1, 2, \cdots, N-1\}$. Here $D_{i}$ denotes the partial derivative of
$\mathbb{L}$ with respect to the $i$th argument. Notice that $t_{k}$ are
parameters here. The reason we still call them arguments is that we want
to keep the notations consistent since in next section they will play the
role of variables.

Now we prove that the discrete flow $\Phi$ preserves a discrete version of the
 Lagrange two form
\begin{align}
\omega_{L}^{v}=d\theta^{v}_{L}, \quad\quad \theta^{v}_{L}=
\frac{\partial L}{\partial \dot{q}}dq. \label{2.a}
\end{align}
We do this directly
from the variational point of view, consistent with the continuous case [\ref{m1}].

As in continuous case, we calculate $d\mathbb{S}$ for variations with varied
endpoints.
\small
\begin{align}
 &d\mathbb{S}(q_{0}, q_{1}, \cdots,  q_{N})\cdot (\delta q_{0}, \delta q_{1},
      \cdots, \delta q_{N})\notag \\
  &=\left.\frac{d}{d\epsilon}\right|_{\epsilon=0}
     \mathbb{S}( q_{0}+\epsilon\delta q_{0}, q_{1}+\epsilon\delta q_{1},
 \cdots,  q_{N}+\epsilon\delta q_{N})\notag \\
  &=\sum_{k=0}^{N-1}(D_{2}\mathbb{L}(t_{k}, q_{k}, t_{k+1}, q_{k+1})\delta q_{k}+
        D_{4}\mathbb{L}(t_{k}, q_{k}, t_{k+1}, q_{k+1})\delta q_{k+1})(t_{k+1}-t_{k})\notag \\
  &=\sum_{k=0}^{N-1}D_{2}\mathbb{L}(t_{k}, q_{k}, t_{k+1}, q_{k+1})(t_{k+1}-t_{k})\delta q_{k}
     +\sum_{k=1}^{N} D_{4}\mathbb{L}(t_{k-1}, q_{k-1}, t_{k}, q_{k})
          (t_{k}-t_{k-1})\delta q_{k}\notag \\
   &=\sum_{k=1}^{N-1}(D_{2}\mathbb{L}(t_{k}, q_{k}, t_{k+1}, q_{k+1})(t_{k+1}-t_{k})
                     +D_{4}\mathbb{L}(t_{k-1}, q_{k-1}, t_{k}, q_{k})
                     (t_{k}-t_{k-1}))\delta q_{k}\notag\\
    &\quad +D_{2}\mathbb{L}(t_{0}, q_{0}, t_{1}, q_{1})(t_{1}-t_{0})\delta q_{0}
           +D_{4}\mathbb{L}(t_{N-1}, q_{N-1}, t_{N}, q_{N})(t_{N}-t_{N-1})
  \delta q_{N}. \label{2.8}
\end{align}
\normalsize

We can see that the last two terms in (\ref{2.8}) come from the boundary variations.
Based on the
boundary variations, we can define two one forms on $M\times M$
\begin{align}
&\theta_{\mathbb{L}}^{v-}(t_{k}, q_{k}, t_{k+1}, q_{k+1})
                    =D_{2}\mathbb{L}(t_{k}, q_{k}, t_{k+1}, q_{k+1})
                (t_{k+1}-t_{k}) dq_{k}, \label{2.9}
\end{align}
and
\begin{align}
 \theta_{\mathbb{L}}^{v+}(t_{k}, q_{k}, t_{k+1}, q_{k+1})
                =D_{4}\mathbb{L}(t_{k}, q_{k}, t_{k+1}, q_{k+1})
                (t_{k+1}-t_{k}) dq_{k+1}. \label{2.10}
\end{align}
Here we have used the notation in [\ref{m1}]. We regard the pair $(\theta_{\mathbb{L}}^{v-},
\theta_{\mathbb{L}}^{v+})$ as being the discrete version of the  Lagrange
one form $\theta^{v}_{L}$  in (\ref{2.a}).

Now we parameterize the solutions of the discrete variational principle by
the initial condition
$( q_{0},  q_{1})$, and  restrict $\mathbb{S}$ to that solution
space.  Then Eq. (\ref{2.8}) becomes
\begin{align}
 d&\mathbb{S}(q_{0}, q_{1}, \cdots,  q_{N})\cdot (\delta q_{0}, \delta q_{1},
      \cdots,  \delta q_{N})\notag\\
 &=\theta_{\mathbb{L}}^{-}(t_{0}, q_{0}, t_{1}, q_{1})\cdot
                        (\delta q_{0},  \delta q_{1})
    +\theta_{\mathbb{L}}^{+}(t_{N-1}, q_{N-1}, t_{N}, q_{N})\cdot
                (\delta q_{N-1}, \delta q_{N})\notag\\
 &=\theta_{\mathbb{L}}^{-}(t_{0}, q_{0}, t_{1}, q_{1})\cdot
                        ( \delta q_{0},   \delta q_{1})
   +(\Phi^{N-1})^{*}\theta_{\mathbb{L}}^{+}(t_{0}, q_{0}, t_{1}, q_{1})\cdot
        ( \delta q_{0}, \delta q_{1}). \label{2.11}
\end{align}
From (\ref{2.11}), we can obtain
\begin{align}
d\mathbb{S}=\theta_{\mathbb{L}}^{v-}+(\Phi^{N-1})^{*}\theta_{\mathbb{L}}^{v+}.
\label{2.12}
\end{align}
The Eq. (\ref{2.12}) holds for arbitrary $N>1$. Taking N=2 leads to
\begin{align}
d\mathbb{S}=\theta_{\mathbb{L}}^{v-}+\Phi^{*}\theta_{\mathbb{L}}^{v+}.
\label{2.13}
\end{align}

By taking exterior differentiation of (\ref{2.13}), we obtain
\begin{align}
  \Phi^{*}(d\theta_{\mathbb{L}}^{v+})=-d\theta_{\mathbb{L}}^{v-}. \label{2.14}
\end{align}
From the definition of $\theta_{\mathbb{L}}^{-}$ and $\theta_{\mathbb{L}}^{+}$, we
know that
\begin{align*}
   \theta_{\mathbb{L}}^{v-}+\theta_{\mathbb{L}}^{v+}=d((t_{k+1}-t_{k})\mathbb{L}),
\end{align*}
which means
$d\theta_{\mathbb{L}}^{v+}=-d\theta_{\mathbb{L}}^{v-}$.

Defining
\begin{align}
 \omega_{\mathbb{L}}^{v}=: d\theta_{\mathbb{L}}^{v+}=
-d\theta_{\mathbb{L}}^{v-}, \label{2.16}
\end{align}
 we  show from (\ref{2.14}) that the discrete flow $\Phi$ preserves the discrete
 Lagrange two form $\omega^{v}_{\mathbb{L}}$:
\begin{align}
  \Phi^{*}(\omega^{v}_{\mathbb{L}})=\omega_{\mathbb{L}}^{v}. \label{2.17}
\end{align}

Let us consider an example. For the classical Lagrangian
\begin{align}
 L(t, q, \,\dot{q})=\frac{1}{2}\dot{q}^{2}-V(q), \label{2.18}
\end{align}
we choose the discrete Lagrangian $\mathbb{L}(t_{k}, q_{k}, t_{k+1}, q_{k+1})$
as
\begin{align}
 \mathbb{L}(t_{k}, q_{k}, t_{k+1}, q_{k+1})=\frac{1}{2}\left(\frac{q_{k+1}-q_{k}}
  {t_{k+1}-t_{k}}\right)^{2}-V\left(\frac{q_{k+1}+q_{k}}{2}\right). \label{2.19}
\end{align}
The discrete Euler-Lagrange equation (\ref{2.3}) becomes
\begin{align}
\left(\frac{q_{k+1}-q_{k}}{t_{k+1}-t_{k}}-\frac{q_{k}-q_{k-1}}{t_{k}-t_{k-1}}
\right)+\frac{V'(\bar{q}_{k})(t_{k+1}-t_{k})+V'(\bar{q}_{k-1})
  (t_{k}-t_{k-1})}{2}=0, \label{2.20}
\end{align}
which preserves the Lagrange two form
\begin{align}
  \left(\frac{1}{t_{k+1}-t_{k}}+\frac{t_{k+1}-t_{k}}{4}V''
   (\bar{q}_{k})\right)dq_{k+1}\wedge dq_{k}, \label{2.21}
\end{align}
where $\bar{q}_{k}=\frac{q_{k}+q_{k+1}}{2}, \quad
  \bar{q}_{k-1}=\frac{q_{k-1}+q_{k}}{2}.$

If we take fixed variables $t_{k+1}-t_{k}=t_{k}-t_{k-1}=h$, then (\ref{2.20})
becomes
\begin{align*}
\frac{q_{k+1}-2q_{k}+q_{k-1}}{h^{2}}+
\frac{V'(\bar{q}_{k})+V'(\bar{q}_{k-1})}{2}=0,
\end{align*}
which preserves the Lagrange two form
\begin{align*}
  \left(\frac{1}{h}+\frac{h}{4}V''
   (\bar{q}_{k})\right)dq_{k+1}\wedge dq_{k}.
\end{align*}
\subsection{Discrete mechanics based on  finite element methods}
Now we consider discrete mechanics based on  finite element methods.
Let us go back to the variation problem of the action factional (\ref{1.1}).
 The finite element method is an approximate method for solving the
variation problem. Instead of solving the variation problem in the
space $\mathcal{C}^{2}([a,b])$, the finite element method solves the problem
in a subspace $V_{h^{m}}([a, b])$ of $\mathcal{C}^{2}([a,b])$. $V_{h^{m}}([a, b])$ consists of
piecewise $m$-degree polynomials interpolating the curves $q(t)\in \mathcal{C}^{2}([a,b])$.

First, we consider the piecewise linear interpolation. Given a partition of [a, b]
\begin{align*}
 a=t_{0}<t_{1}<\cdots<t_{k}<\cdots<t_{N-1}<t_{N}=b,
\end{align*}
the intervals $I_{k}=[t_{k}, t_{k+1}]$ are called elements. $h_{k}=t_{k+1}-t_{k}$.
$V_{h}([a, b])$ consists of piecewise linear function interpolating $q(t)$
at $(t_{k}, q_{k}), \,\, k=0, 1, \cdots, N$. Now we derive the expressions
of $q_{h}(t)\in V_{h}([a, b])$. First we construct the basis functions
 $\varphi_{k}(t)$,  which are piecewise linear function on $[a, b]$
satisfying $\varphi_{k}(t_{i})=\delta ^{i}_{k}, \,\, i, k=0, 1, \cdots, N$.
\begin{align}
 \varphi_{0}(t)=\begin{cases}
                 1-\frac{t-t_{0}}{h_{0}},  & t_{0}\leqslant t\leqslant t_{1};\\
                ~~ 0,          &\text{otherwise};
                \end{cases}
\quad
\varphi_{N}(t)=\begin{cases}
                 1+\frac{t-t_{N}}{h_{N-1}},  & t_{N-1}\leqslant t\leqslant t_{N};\\
                 ~~ 0,          &\text{otherwise};
                \end{cases} \label{2.24}
\end{align}
and for $k=1, 2, \cdots, N-1$,
\begin{align}
 \varphi_{k}(t)=\begin{cases}
         1+\frac{t-t_{k}}{h_{k-1}},  & t_{k-1}\leqslant t\leqslant t_{k};\\
         1-\frac{t-t_{k}}{h_{k}},  & t_{k}\leqslant t\leqslant t_{k+1};\\
         ~~ 0,          &\text{otherwise}.
                \end{cases} \label{2.25}
\end{align}
Using these basis functions, we obtain the expression $q_{h}\in V_{h}([a, b])$:
\begin{align}
 q_{h}(t)=\sum_{k=0}^{N}q_{k}\varphi_{k}(t).
\end{align}
In the space $V_{h}([a, b])$, the action functional (\ref{1.1}) becomes
\begin{align}
    S((t, q_{h}(t)))&=\int_{a}^{b}L(t, q_{h}(t), \dot{q}_{h}(t))\,dt\notag \\
     &=\sum_{k=0}^{N-1}\int_{t_{k}}^{t_{k+1}}
   L\left(t, \sum_{i=0}^{N}(q_{i}\varphi_{i}(t)), \frac{d}{dt}
        \sum_{i=0}^{N}(q_{i}\varphi_{i}(t))\right)dt \notag\\
     &=\sum_{k=0}^{N-1}\mathbb{L}(t_{k}, q_{k}, t_{k+1}, q_{k+1})
   (t_{k+1}-t_{k}), \label{2.26}
\end{align}
where
\begin{align}
   \mathbb{L}(t_{k}, q_{k}, t_{k+1}, q_{k+1})
 =&\frac{1}{t_{k+1}-t_{k}}\int_{t_{k}}^{t_{k+1}}
   L\left(t, \sum_{i=0}^{N}(q_{i}\varphi_{i}(t)), \frac{d}{dt}
        \sum_{i=0}^{N}(q_{i}\varphi_{i}(t))\right)dt\notag\\
  =&\frac{1}{t_{k+1}-t_{k}}\int_{t_{k}}^{t_{k+1}}
   L\left(t, \sum_{i=k}^{k+1}(q_{i}\varphi_{i}(t)), \frac{d}{dt}
        \sum_{i=k}^{k+1}(q_{i}\varphi_{i}(t))\right)dt.\label{2.27}
\end{align}

Therefore, restricting to the subspace $V_{h}([a,
b])$ of $\mathcal{C}^{2}([a, b])$, the original
variational problem reduces to the extremum
problem of the function (\ref{2.26}) in $q_{k},\,
k=0, 1, \cdots, N$. Notice that (\ref{2.26}) is
just one of the discrete actions (\ref{2.1}).
Thus, what remains to be done is just to perform
the same calculations on (\ref{2.26}) as on
(\ref{2.1}). We can then obtain the discrete
Euler-Lagrange equation (\ref{2.3}) that
preserves the discrete Lagrange two form
(\ref{2.16}). Therefore, discrete mechanics based
on finite element methods consists of two steps:
first, use finite element methods to obtain a
kind of discrete Lagrangian; second, use the
method of Veselov mechanics to obtain the
variational integrators.

Let us consider the previous example again.  For the classical Lagrangian
(\ref{2.18}), we choose the discrete Lagrangian
$\mathbb{L}(t_{k}, q_{k}, t_{k+1}, q_{k+1})$ as
\begin{align}
 &\mathbb{L}(t_{k}, q_{k}, t_{k+1}, q_{k+1})\notag\\
 &\quad =\frac{1}{t_{k+1}-t_{k}}\int_{t_{k}}^{t_{k+1}}
   \left(\frac{1}{2} \left(\frac{d}{dt}
    \sum_{i=0}^{N}(q_{i}\varphi_{i}(t))\right)^2-V\left(
      \sum_{i=0}^{N}(q_{i}\varphi_{i}(t))\right)\right) dt \notag\\
  &\quad=\frac{1}{t_{k+1}-t_{k}}\int_{t_{k}}^{t_{k+1}}
   \left(\frac{1}{2} \left(\frac{q_{k+1}-q_{k}}{t_{k+1}-t_{k}}
    \right)^2-V\left(\frac{t_{k+1}-t}{t_{k+1}-t_{k}}q_{k}
  +\frac{t-t_{k}}{t_{k+1}-t_{k}}q_{k+1}\right)\right) dt \notag\\
  &\quad=\frac{1}{2} \left(\frac{q_{k+1}-q_{k}}{t_{k+1}-t_{k}}
    \right)^2-F(q_{k}, q_{k+1}), \label{2.28}
\end{align}
where
\begin{align}
 F(q_{k}, q_{k+1})=\frac{1}{t_{k+1}-t_{k}}\int_{t_{k}}^{t_{k+1}}
       V\left(\frac{t_{k+1}-t}{t_{k+1}-t_{k}}q_{k}+
   \frac{t-t_{k}}{t_{k+1}-t_{k}}q_{k+1}\right) dt. \label{2.29}
\end{align}
The discrete Euler-Lagrange equation (\ref{2.3}) becomes
\small
\begin{align}
\left(\frac{q_{k+1}-q_{k}}{t_{k+1}-t_{k}}-\frac{q_{k}-q_{k-1}}{t_{k}-t_{k-1}}
\right)+\frac{\partial F(q_{k},
q_{k+1})}{\partial q_{k}}(t_{k+1}-t_{k})
+\frac{\partial F(q_{k-1}, q_{k})}{\partial
q_{k}}(t_{k}-t_{k-1}) =0, \label{2.30}
\end{align}
\normalsize
which preserves the Lagrange two form
\begin{align}
  \left(\frac{1}{t_{k+1}-t_{k}}+(t_{k+1}-t_{k})
  \frac{\partial^{2}F(q_{k}, q_{k+1})}{\partial q_{k}\partial q_{k+1}}
   \right)dq_{k+1}\wedge dq_{k}. \label{2.31}
\end{align}
Again, if we take fixed time steps $t_{k+1}-t_{k}=t_{k}-t_{k-1}=h$, (\ref{2.30})
becomes
\begin{align*}
\frac{q_{k+1}-2q_{k}+q_{k-1}}{h^{2}}
+\frac{\partial F(q_{k}, q_{k+1})}{\partial
q_{k}} +\frac{\partial F(q_{k-1},
q_{k})}{\partial q_{k}} =0,
\end{align*}
which preserves the Lagrange two form
\begin{align*}
  \left(\frac{1}{h}+h
  \frac{\partial^{2}F(q_{k}, q_{k+1})}{\partial q_{k}\partial q_{k+1}}
   \right)dq_{k+1}\wedge dq_{k}.
\end{align*}

Suppose $q_{k}$ is the solution of (\ref{2.30}) and $q(t)$ is the solution of
\begin{align}
  \frac{d^{2}q}{dt^{2}}+\frac{\partial V(q)}{\partial q}=0,\label{2.32}
\end{align}
then from the convergence theory of finite element methods [\ref{c2}, \ref{l4}], we have
 \begin{align}
\|q(t)-q_{h}(t)\|\leqslant Ch^{2}, \label{2.33}
\end{align}
where $\|\cdot\|$ is the $L^{2}$ norm. $q_{h}(t)=\sum_{k=0}^{N}q_{k}$,
$h=\underset{k}{\text{max}}\{h_{k}\}$ and $C$ is a constant independent of $h$.

If we use mid-point numerical integration formula in (\ref{2.29}), we obtain
\begin{align*}
 F(q_{k}, q_{k+1})=&\frac{1}{t_{k+1}-t_{k}}\int_{t_{k}}^{t_{k+1}}
       V\left(\frac{t_{k+1}-t}{t_{k+1}-t_{k}}q_{k}+
   \frac{t-t_{k}}{t_{k+1}-t_{k}}q_{k+1}\right) dt \\
  \approx &V\left(\frac{q_{k}+q_{k+1}}{2}\right).
\end{align*}
In this case, (\ref{2.30}) is just (\ref{2.20}).
We can also use trapezoid formula or Simpson
formula and so on to integrate (\ref{2.29})
numerically  and obtain another  kind of discrete
Lagrangian.

\subsection{Discrete mechanics with Lagrangian of high order}

Now we consider piecewise quadratic polynomial
interpolation, which will result in a kind of
discrete Lagrangian of high order. To this aim,
we add an auxiliary node $t_{k+\frac{1}{2}}$ to
each element $I_{k}=[t_{k}, t_{k+1}]$. There are
two kinds of quadratic basis functions:
$\phi_{k}(t)$ for nodes $t_{k}$ and
$\phi_{k+\frac{1}{2}}(t)$ for $t_{k+\frac{1}{2}}$
that satisfy
\begin{align*}
        &\phi_{k}(t_{i})=\delta_{i}^{k}, \quad \phi_{k}(t_{i+\frac{1}{2}})=0,\\
       &\phi_{k+\frac{1}{2}}(t_{i+\frac{1}{2}})=\delta_{i}^{k}, \quad
       \phi_{k+\frac{1}{2}}(t_{i})=0, \quad i, k=0, 1, \cdots, N.
\end{align*}
 We list the basis functions as follows:
\begin{align}
 \phi_{0}(t)=\begin{cases}
  \left(\frac{2(t-t_{0})}{h_{0}}-1\right)\!\!\left(
     \frac{t-t_{0}}{h_{0}}-1\right),
  & t_{0}\leqslant t\leqslant t_{1};\\
                ~~ 0,          &\text{otherwise};
                \end{cases}\label{2.34}
\end{align}
\begin{align}
\phi_{N}(t)=\begin{cases}
    \left(\frac{2(t_{N}-t)}{h_{N-1}}-1\right)\!\!\left(
     \frac{t_{N}-t}{h_{N-1}}-1\right),
  & t_{N-1}\leqslant t\leqslant t_{N};\\
                 ~~0,          &\text{otherwise};
                \end{cases} \label{2.35}
\end{align}
and for $k=1, 2, \cdots, N-1$,
\begin{align}
 \phi_{k}(t)=\begin{cases}
    \left(\frac{2(t_{k}-t)}{h_{k-1}}-1\right)\!\!\left(
     \frac{t_{k}-t}{h_{k-1}}-1\right),
  & t_{k-1}\leqslant t\leqslant t_{k};\\
  \left(\frac{2(t-t_{k})}{h_{k}}-1\right)\!\!\left(
     \frac{t-t_{k}}{h_{k}}-1\right),
  & t_{k}\leqslant t\leqslant t_{k+1};\\
         ~~0,          &\text{otherwise};
                \end{cases} \label{2.36}
\end{align}
and for $k=0, 1, \cdots, N-1$,
\begin{align}
 \phi_{k+\frac{1}{2}}(t)=\begin{cases}
    4\frac{t-t_{k}}{h_{k}}\!\left(1-
     \frac{t-t_{k}}{h_{k}}\right),
  & t_{k}\leqslant t\leqslant t_{k+1};\\
        ~~ 0,          &\text{otherwise}.
                \end{cases} \label{2.37}
\end{align}
Using these basis functions, we construct  subspace $V_{h^{2}}([a, b]) $
of  $\mathcal{C}^{2}([a, b])$:
\begin{align}
  q_{h^{2}}(t)=\sum_{k=0}^{N}q_{k}\phi_{k}(t)+\sum_{k=0}^{N-1}q_{k+\frac{1}{2}}
\phi_{k+\frac{1}{2}}(t), \quad \quad q_{h^{2}}(t)\in V_{h^{2}}([a, b]). \label{2.38b}
\end{align}
In the space $V_{h^{2}}([a, b])$, the action functional (\ref{1.1}) becomes
\begin{align}
    S((t, q_{h^{2}}(t)))&=\int_{a}^{b}L(t, q_{h^{2}}(t), \dot{q}_{h^{2}}(t))\,dt\notag \\
     &=\sum_{k=0}^{N-1}\int_{t_{k}}^{t_{k+1}}
    L(t, q_{h^{2}}(t), \dot{q}_{h^{2}}(t))\,dt\notag\\
     &=\sum_{k=0}^{N-1}\mathbb{L}(t_{k}, q_{k}, q_{k+\frac{1}{2}}, t_{k+1}, q_{k+1})
   (t_{k+1}-t_{k}), \label{2.38}
\end{align}
where
\begin{align}
   \mathbb{L}(t_{k}, q_{k}, q_{k+\frac{1}{2}}, t_{k+1}, q_{k+1})
 =\frac{1}{t_{k+1}-t_{k}}\int_{t_{k}}^{t_{k+1}}
   L(t, q_{h^{2}}(t), \dot{q}_{h^{2}}(t))\,dt.\label{2.39b}
\end{align}
For the discrete action (\ref{2.38}), we have
\small
\begin{align}
 &d\mathbb{S}(q_{0}, q_{\frac{1}{2}}, q_{1}, \cdots, q_{N-1+\frac{1}{2}},
   q_{N})\cdot (\delta q_{0}, \delta q_{\frac{1}{2}},  \delta q_{1},
      \cdots, \delta q_{N-1+\frac{1}{2}}, \delta q_{N})\notag \\
  &=\sum_{k=0}^{N-1}(D_{2}\mathbb{L}(\w_{k})\delta q_{k}+
         D_{3}\mathbb{L}(\w_{k})\delta q_{k+\frac{1}{2}}
        +D_{5}\mathbb{L}(\w_{k})\delta q_{k+1})(t_{k+1}-t_{k})\notag \\
  &=\sum_{k=0}^{N-1}(D_{2}\mathbb{L}(\w_{k})\delta q_{k}
      +D_{3}\mathbb{L}(\w_{k})\delta q_{k+\frac{1}{2}})(t_{k+1}-t_{k})
     +\sum_{k=1}^{N} D_{5}\mathbb{L}(\w_{k-1})
          (t_{k}-t_{k-1})\delta q_{k}\notag
          \\[2mm]
   &=\sum_{k=1}^{N-1}(D_{2}\mathbb{L}(\w_{k})(t_{k+1}-t_{k})
                     +D_{5}\mathbb{L}(\w_{k-1})
                     (t_{k}-t_{k-1}))\delta q_{k}+\sum_{k=0}^{N-1}
  D_{3}\mathbb{L}(\w_{k})\delta q_{k+\frac{1}{2}}(t_{k+1}-t_{k})\notag\\
    &\quad +D_{2}\mathbb{L}(\w_{0})(t_{1}-t_{0})\delta q_{0}
           +D_{5}\mathbb{L}(\w_{N-1})(t_{N}-t_{N-1})\delta q_{N}, \label{2.39}
\end{align}
\normalsize
where $\w_{k}=(t_{k}, q_{k}, q_{k+\frac{1}{2}}, t_{k+1}, q_{k+1}), \,\, k=0, 1,
\cdots, N-1$.
From (\ref{2.39}), we obtain the discrete Euler-Lagrange equation
\begin{align}
 &D_{2}\mathbb{L}(\w_{k})(t_{k+1}-t_{k})+D_{5}\mathbb{L}(\w_{k-1})
                     (t_{k}-t_{k-1})=0,\label{2.40}\\
 &D_{3}\mathbb{L}(t_{k}, q_{k}, q_{k+\frac{1}{2}}, t_{k+1}, q_{k+1})=0,\label{2.41}\\
 &D_{3}\mathbb{L}(t_{k-1}, q_{k-1}, q_{k-1+\frac{1}{2}}, t_{k}, q_{k})=0. \label{2.42}
\end{align}

We can from (\ref{2.41}) and (\ref{2.42}) solve for $q_{k+\frac{1}{2}}$
and $q_{k-1+\frac{1}{2}}$ respectively, then substitute them into (\ref{2.40}) and
finally solve for $q_{k+1}$. Therefore, the discrete Euler-Lagrange equation
(\ref{2.40}-\ref{2.42}) still determines a discrete flow
\begin{align*}
   &\Psi:\,M\times M\to M\times M \\
   &\Psi(t_{k-1}, q_{k-1}, t_{k}, q_{k})=(t_{k}, q_{k},t_{k+1}, q_{k+1}).
\end{align*}
From (\ref{2.39}), we can define two one forms
\begin{align*}
&\Theta_{\mathbb{L}}^{v-}(t_{k}, q_{k}, q_{k+\frac{1}{2}}, t_{k+1}, q_{k+1})
                    =D_{2}\mathbb{L}(t_{k}, q_{k}, q_{k+\frac{1}{2}}, t_{k+1}, q_{k+1})
                (t_{k+1}-t_{k}) dq_{k},
\end{align*}
and
\begin{align*}
 \Theta_{\mathbb{L}}^{v+}(t_{k}, q_{k}, q_{k+\frac{1}{2}}, t_{k+1}, q_{k+1})
                =D_{5}\mathbb{L}(t_{k}, q_{k}, q_{k+\frac{1}{2}}, t_{k+1}, q_{k+1})
                (t_{k+1}-t_{k}) dq_{k+1}.
\end{align*}
Using the same method as before, we can prove that
\begin{align}
   \Psi^{*}(d\Theta_{\mathbb{L}}^{v+})=-d\Theta_{\mathbb{L}}^{v-}. \label{2.43}
\end{align}
From the definition of $\Theta_{\mathbb{L}}^{v-}$ and $\Theta_{\mathbb{L}}^{v+}$,
we have
\begin{align}
  \Theta_{\mathbb{L}}^{v-}+\Theta_{\mathbb{L}}^{v+}
=d((t_{k+1}-t_{k})L)-D_{3}L(t_{k}, q_{k}, q_{k+\frac{1}{2}}, t_{k+1}, q_{k+1})
dq_{k+\frac{1}{2}}. \label{2.44}
\end{align}
From (\ref{2.41}), we obtain
$D_{3}L(t_{k}, q_{k}, q_{k+\frac{1}{2}}, t_{k+1}, q_{k+1})=0$. Thus,
\begin{align*}
  \Theta_{\mathbb{L}}^{v-}+\Theta_{\mathbb{L}}^{v+}
=d((t_{k+1}-t_{k})L),
\end{align*}
which means
 \begin{align}
  d\Theta_{\mathbb{L}}^{v+}=-d\Theta_{\mathbb{L}}^{v-}.\label{2.45}
 \end{align}
From (\ref{2.43}) and (\ref{2.45}), we arrive at
\begin{align}
 \Psi^{*}(\Omega_{\mathbb{L}}^{v})= \Omega_{\mathbb{L}}^{v}, \label{2.46}
\end{align}
where $ \Omega_{\mathbb{L}}^{v}=d \Theta_{\mathbb{L}}^{v+}$.

For the classical Lagrangian
(\ref{2.18}),  From (\ref{2.38b}) and (\ref{2.39b}), we obtain
\begin{align}
   &\mathbb{L}(t_{k}, q_{k}, q_{k+\frac{1}{2}}, t_{k+1}, q_{k+1})\notag\\
 &\quad =\frac{1}{t_{k+1}-t_{k}}\int_{t_{k}}^{t_{k+1}}
   \left(\frac{1}{2}(\dot{q}_{h^{2}}(t))^{2}-V(q_{h^{2}}(t))\right)dt\notag\\
 &\quad =\frac{1}{2}\left(\frac{1}{3}a^{2}(t_{k+1}^{2}+t_{k}t_{k+1}+t_{k}^{2})
 +ab(t_{k}+t_{k+1})+b^{2}\right)\notag\\
 &\quad \quad -G(q_{k}, q_{k+\frac{1}{2}}, q_{k+1}),
\label{2.47}
\end{align}
where
\begin{align*}
   &a=\frac{4}{h_{k}^{2}}\left(q_{k}+q_{k+1}-2q_{k+\frac{1}{2}}\right), \\
   &b=\frac{1}{h_{k}^{2}}\left(4(t_{k}+t_{k+1})q_{k+\frac{1}{2}}
      -(3t_{k}+t_{k+1})q_{k+1} -(t_{k}+3t_{k+1})q_{k}\right),
\end{align*}
and
\begin{align*}
   G(q_{k}, q_{k+\frac{1}{2}}, q_{k+1})
    =\frac{1}{t_{k+1}-t_{k}}\int_{t_{k}}^{t_{k+1}}
     V\left(q_{k}f_{k}(t)+q_{k+1}f_{k+1}(t)+q_{k+\frac{1}{2}}f_{k+\frac{1}{2}}(t)
      \right)dt,
\end{align*}
where
\begin{align*}
      &f_{k}(t)=\left(\frac{2(t-t_{k})}{h_{k}}-1\right)\!\!\left(
     \frac{t-t_{k}}{h_{k}}-1\right),\\
      &f_{k+1}(t)=\left(\frac{2(t_{k+1}-t)}{h_{k}}
      -1\right)\!\!\left(
     \frac{t_{k+1}-t}{h_{k}}-1\right),\\
      &f_{k+\frac{1}{2}}(t)=4\frac{t-t_{k}}{h_{k}}\!\left(1-
     \frac{t-t_{k}}{h_{k}}\right).
\end{align*}
For the discrete Lagrangian (\ref{2.47}), the discrete Euler-Lagrangian
equation (\ref{2.40}-\ref{2.42}) becomes
\begin{align}
&a_{1}q_{k-1}+a_{2}q_{k}+a_{3}q_{k+1}+a_{4}q_{k-\frac{1}{2}}
 +a_{4}q_{k+\frac{1}{2}}-d_{1}h_{k}-d_{2}h_{k-1}=0,\label{2.48}\\
&-\frac{8}{3h_{k}^{2}}\left(q_{k}+q_{k+1}-2q_{k+\frac{1}{2}}\right)
-\frac{\partial G(q_{k}, q_{k+\frac{1}{2}}, q_{k+1})}
{\partial q_{k+\frac{1}{2}}}=0,\label{2.49}\\
&-\frac{8}{3h_{k-1}^{2}}\left(q_{k-1}+q_{k}-2q_{k-1+\frac{1}{2}}\right)
-\frac{\partial G(q_{k-1}, q_{k-1+\frac{1}{2}}, q_{k})}
{\partial q_{k-1+\frac{1}{2}}}=0, \label{2.50}
\end{align}
where
\begin{align*}
     &a_{1}=\frac{1}{3}\frac{1}{h_{k-1}}, \quad
   a_{2}=\frac{7}{3}\left(\frac{1}{h_{k-1}}+\frac{1}{h_{k}}\right),\quad
   a_{3}=\frac{1}{3}\frac{1}{h_{k}},\\
    &a_{4}=-\frac{8}{3}\frac{1}{h_{k-1}},\quad
     a_{5}=-\frac{8}{3}\frac{1}{h_{k}},\\
   &d_{1}=\frac{\partial G(q_{k}, q_{k+\frac{1}{2}}, q_{k+1})}
  {\partial q_{k}}, \quad
     d_{2}=\frac{\partial G(q_{k-1}, q_{k-1+\frac{1}{2}}, q_{k})}
{\partial q_{k}}.
\end{align*}
The solution of (\ref{2.48}-\ref{2.50}) preserves the Lagrange two form
\begin{align}
\left(\frac{1}{3h_{k}}-h_{k}\frac{\partial^{2}G(q_{k}, q_{k+\frac{1}{2}},
q_{k+1})}{\partial q_{k}\partial q_{k+1}}-M\right)dq_{k}\wedge dq_{k+1},
\label{2.51}
\end{align}
where
\[
  M=\frac{\left(\frac{16}{3h_{k}}+h_{k}\frac{\partial^{2}G(q_{k}, q_{k+\frac{1}{2}},
q_{k+1})}{\partial q_{k+\frac{1}{2}}\partial q_{k}}\right)
\left(\frac{16}{3h_{k}}+h_{k}\frac{\partial^{2}G(q_{k}, q_{k+\frac{1}{2}},
q_{k+1})}{\partial q_{k+\frac{1}{2}}\partial q_{k}}\right)}
{\left(\frac{32}{3h_{k}}-h_{k}\frac{\partial^{2}G(q_{k}, q_{k+\frac{1}{2}},
q_{k+1})}{\partial q^{2}_{k+\frac{1}{2}}}\right)}.
\]
If we take the fixed time steps
$h_{k-1}=h_{k}=h$, then (\ref{2.48}-\ref{2.50}))
become
\begin{align}
&\frac{q_{k-1}-8q_{k-\frac{1}{2}}+14q_{k}-8q_{k+\frac{1}{2}}+q_{k+1}}{3h^{2}}
 -d_{1}h_{k}-d_{2}h_{k-1}=0,\label{2.52}\\
&-\frac{8}{3h^{2}}\left(q_{k}+q_{k+1}-2q_{k+\frac{1}{2}}\right)
-\frac{\partial G(q_{k}, q_{k+\frac{1}{2}}, q_{k+1})}
{\partial q_{k+\frac{1}{2}}}=0,\label{2.53}\\
&-\frac{8}{3h^{2}}\left(q_{k-1}+q_{k}-2q_{k-1+\frac{1}{2}}\right)
-\frac{\partial G(q_{k-1}, q_{k-1+\frac{1}{2}}, q_{k})}
{\partial q_{k-1+\frac{1}{2}}}=0, \label{2.54}
\end{align}
which preserve
\begin{align}
\left(\frac{1}{3h}-h\frac{\partial^{2}G(q_{k}, q_{k+\frac{1}{2}},
q_{k+1})}{\partial q_{k}\partial q_{k+1}}-M\right)dq_{k}\wedge dq_{k+1},
\label{2.55}
\end{align}
where
\[
  M=\frac{\left(\frac{16}{3h_{k}}+h\frac{\partial^{2}G(q_{k}, q_{k+\frac{1}{2}},
q_{k+1})}{\partial q_{k+\frac{1}{2}}\partial q_{k}}\right)
\left(\frac{16}{3h}+h\frac{\partial^{2}G(q_{k}, q_{k+\frac{1}{2}},
q_{k+1})}{\partial q_{k+\frac{1}{2}}\partial q_{k}}\right)}
{\left(\frac{32}{3h}-h\frac{\partial^{2}G(q_{k}, q_{k+\frac{1}{2}},
q_{k+1})}{\partial q^{2}_{k+\frac{1}{2}}}\right)}.
\]

Suppose $q_{k}$ is the solution of (\ref{2.48}-\ref{2.50})) and
$q(t)$ is the solution of (\ref{2.32}),
then from the convergence theory of finite element methods [\ref{c2}, \ref{l4}], we have
 \begin{align}
\|q(t)-q_{h^{2}}(t)\|\leqslant Ch^{3}, \label{2.56}
\end{align}
where
\[
 q_{h^{2}}(t)=\sum_{k=0}^{N}q_{k}\phi_{k}(t)+\sum_{k=0}^{N-1}q_{k+\frac{1}{2}}
\phi_{k+\frac{1}{2}}(t),
\]
 $h=\underset{k}{\text{max}}\{h_{k}\}$ and $C$ is a constant independent of $h$.


\sect{Time steps as variables} In \S 2, the time
steps $t_{k}$ play the role of parameters. They
are determined beforehand according to some
requirements. In fact, we can also regard $t_{k}$
as variables and the variation of the discrete
action with respect to $t_{k}$ yields the
discrete energy conservation law. This fact was
first observed by Lee [ \ref{l1}, \ref{l2},
\ref{l3}]. The symplecticity of the resulting
integrators was investigated in [\ref{k1},
\ref{c1}]. These results
 also apply to the discrete mechanics based on finite element methods.

We regard $t_{k}$ as variables and  calculate the
variation of the discrete action (\ref{2.1}) as
follows
\begin{align}
 &d\mathbb{S}(t_{0}, q_{0}, \cdots, t_{N}, q_{N})\cdot (\delta t_{0}, \delta q_{0},
      \cdots, \delta t_{N}, \delta q_{N})\notag \\
  &=\left.\frac{d}{d\epsilon}\right|_{\epsilon=0}
     \mathbb{S}(t_{0}+\epsilon\delta t_{0}, q_{0}+\epsilon\delta q_{0},
 \cdots, t_{N}+\epsilon\delta t_{N}, q_{N}+\epsilon\delta q_{N})\notag \\
   &=\sum_{k=1}^{N-1}\left[D_{2}\mathbb{L}(\w_{k})(t_{k+1}-t_{k})
                     +D_{4}\mathbb{L}(\w_{k-1})
                     (t_{k}-t_{k-1})\right]\delta q_{k}\notag\\
    &\quad +\sum_{k=1}^{N-1}\left[D_{1}\mathbb{L}(\w_{k})(t_{k+1}-t_{k})
                     +D_{3}L(\w_{k-1})
                     (t_{k}-t_{k-1})+L(\w_{k-1})
      -L(\w_{k})\right]\delta t_{k}\notag\\
    &\quad +D_{2}L(\w_{0})(t_{1}-t_{0})\delta q_{0}
           +D_{4}L(\w_{N-1})(t_{N}-t_{N-1})\delta q_{N}\notag\\
   &\quad  +\left[D_{1}L(\w_{0})(t_{1}-t_{0})
               -L(\w_{0})\right]\delta t_{0}\notag\\
    &\quad  +\left[D_{3}L(\w_{N-1})(t_{N}-t_{N-1})
              +L(\w_{N-1})\right]\delta t_{N}, \label{3.1}
\end{align}
where $\w_{k}=(t_{k}, q_{k}, t_{k+1}, q_{k+1}), \quad k=0, 1, \cdots, N-1$.
From (\ref{3.1}), we see that the variation $\delta q_{k}$ yields the
discrete Euler-Lagrange equation
\begin{align}
 D_{2}\mathbb{L}(\w_{k})(t_{k+1}-t_{k})
                     +D_{4}\mathbb{L}(\w_{k-1})
                     (t_{k}-t_{k-1})=0 \label{3.2}
\end {align}
and the variation $\delta t_{k}$ yields the discrete energy evolution equation
\begin{align}
D_{1}\mathbb{L}(\w_{k})(t_{k+1}-t_{k})
                     +D_{3}L(\w_{k-1})
                     (t_{k}-t_{k-1})+L(\w_{k-1})
      -L(\w_{k})=0, \label{3.3}
\end{align}
which is a discrete version of (\ref{1.9}). For a conservative $L$, (\ref{3.3})
becomes the discrete energy conservation law.

From the boundary terms in (\ref{3.1}), we can define two one-forms
\begin{align}
\theta_{\mathbb{L}}^{-}(\w_{k})
                    =(D_{1}\mathbb{L}(\w_{k})
            (t_{k+1}-t_{k})-\mathbb{L}(\w_{k}))dt_{k}
            +D_{2}\mathbb{L}(\w_{k})
                (t_{k+1}-t_{k}) dq_{k}, \label{3.4}
\end{align}
and
\begin{align}
 \theta_{\mathbb{L}}^{+}(\w_{k})
                 =(D_{3}\mathbb{L}(\w_{k})
            (t_{k+1}-t_{k})+\mathbb{L}(\w_{k}))dt_{k+1}
            +D_{4}\mathbb{L}(\w_{k})
                (t_{k+1}-t_{k}) dq_{k+1}. \label{3.5}
\end{align}
These two one forms are the discrete version of the extended Lagrange one form
 (\ref{1.12}).

Unlike the continuous case, the solution of (\ref{3.2}) does not satisfy (\ref{3.3})
in general. Therefore, we must solve (\ref{3.2}) and (\ref{3.3}) simultaneously. Using
the same method in \S 2, we can show that the coupled integrator
\begin{align}
\begin{split}
&D_{2}\mathbb{L}(\w_{k})(t_{k+1}-t_{k})
                     +D_{4}\mathbb{L}(\w_{k-1})
                     (t_{k}-t_{k-1})=0,\\
&D_{1}\mathbb{L}(\w_{k})(t_{k+1}-t_{k})
                     +D_{3}L(\w_{k-1})
                     (t_{k}-t_{k-1})+L(\w_{k-1})
      -L(\w_{k})=0 \label{3.6}
\end{split}
\end{align}
preserves the extended Lagrange two form $\omega_{\mathbb{L}}=d\theta_{\mathbb{L}}^{+}$.

For the discrete Lagrangian (\ref{2.28}), (\ref{3.6}) becomes
\small
\begin{align*}
&\left(\frac{q_{k+1}-q_{k}}{t_{k+1}-t_{k}}-\frac{q_{k}-q_{k-1}}{t_{k}-t_{k-1}}
\right)+\frac{\partial F(\w_{k})}{\partial q_{k}}h_{k}
+\frac{\partial F(\w_{k-1})}{\partial q_{k}}h_{k-1}
=0,\\
&\frac{1}{2}\left(\frac{q_{k+1}-q_{k}}{t_{k+1}-t_{k}}\right)^{2}+F(\w_{k})
-\frac{\partial F(\w_{k})}{\partial t_{k}}h_{k}
=\frac{1}{2}\left(\frac{q_{k}-q_{k-1}}{t_{k}-t_{k-1}}\right)^{2}+F(\w_{k-1})
+\frac{\partial F(\w_{k-1})}{\partial t_{k}}h_{k-1}.
\end{align*}
\normalsize For the  kind of high order discrete
Lagrangian, we can obtain similar formulas.


\sect{Conclusions} In this paper, we have
presented the discrete mechanics based on finite
element methods. The finite element methods
provide a natural way to obtain a kind of
discrete Lagrangian. Using piecewise quadratic
functions, we have constructed a high order
discrete Lagrangian and derived the corresponding
variational integrator (the discrete
Euler-Lagrange equation). The symplecticity of
the high order variational integrator has also
been proved.

The time steps can play the role of parameters or the role of variables.
We can obtain discrete energy conservation law when the time steps are
regarded as variables.

Finite element methods are well-established  in
numerical mathematics. Many results of finite
element methods can be applied to the discrete
mechanics presented in this paper such as
convergence theory, error estimates as well as
the solving method of the variational
integrators.

Recently, it has been proved [\ref{g2}] that the
simplectic structure preserves not only on the
phase flow but also on the flows with respect to
symplectic vector fields as long as certain
cohomological condition is satisfied in both
continuous and discrete cases. This should be
able to extend to the cases in this paper. We
will study this issue elsewhere.

\vskip 8mm

\centerline{\bf Acknowledgments} \vskip 2mm

 This work was
supported in part by the National Natural Science
Foundation of China (grant Nos. 90103004,
10171096) and the National Key Project  for Basic
Research of China (G1998030601).



\end{document}